\documentstyle[twoside,multicol,psfig]{article}

\catcode`\@=11
\long\def\@makefntext#1{
\protect\noindent \hbox to 3.2pt {\hskip-.9pt  
$^{{\footnotesize\@thefnmark}}$\hfil}#1\hfill}          %CAN BE USED 

\def\@makefnmark{\hbox to 0pt{$^{\@thefnmark}$\hss}}    %ORIGINAL 
        
\def\ps@myheadings{%
    \let\@oddfoot\@empty\let\@evenfoot\@empty
    \def\@evenhead{\footnotesize\it\leftmark\hfil}%      %EVEN PAGE
    \def\@oddhead{\hfil{\footnotesize\it\rightmark}}%    %ODD PAGE
    \let\@mkboth\@gobbletwo
    \let\sectionmark\@gobble
    \let\subsectionmark\@gobble
    }

\oddsidemargin=\evensidemargin
\newcounter{sectionc}\newcounter{subsectionc}\newcounter{subsubsectionc}
\renewcommand{\section}[1] {\vspace{14pt}\addtocounter{sectionc}{1}
\setcounter{subsectionc}{0}\setcounter{subsubsectionc}{0}\noindent 
        {\bf\thesectionc. #1}\par\vspace{8pt}}
\renewcommand{\subsection}[1] {\vspace{14pt}\addtocounter{subsectionc}{1}
   \setcounter{subsubsectionc}{0}\noindent 
   {\bf\thesectionc.\thesubsectionc. {\kern1pt \bfit #1}}\par\vspace{8pt}}
\renewcommand{\subsubsection}[1] {\vspace{14pt}
    \addtocounter{subsubsectionc}{1}
        \noindent{\thesectionc.\thesubsectionc.\thesubsubsectionc.
        {\kern1pt \it #1}}\par\vspace{8pt}}

\newcounter{appendixc}
\newcounter{subappendixc}[appendixc]
\newcounter{subsubappendixc}[subappendixc]
\renewcommand{\thesubappendixc}{\Alph{appendixc}.\arabic{subappendixc}}
\renewcommand{\thesubsubappendixc}
        {\Alph{appendixc}.\arabic{subappendixc}.\arabic{subsubappendixc}}

\renewcommand{\appendix}[1] {\vspace{14pt}
        \refstepcounter{appendixc}
        \setcounter{figure}{0}
        \setcounter{table}{0}
        \setcounter{lemma}{0}
        \setcounter{theorem}{0}
        \setcounter{corollary}{0}
        \setcounter{definition}{0}
        \setcounter{equation}{0}
        \renewcommand{\thefigure}{\Alph{appendixc}.\arabic{figure}}
        \renewcommand{\thetable}{\Alph{appendixc}.\arabic{table}}
        \renewcommand{\theappendixc}{\Alph{appendixc}}
        \renewcommand{\thelemma}{\Alph{appendixc}.\arabic{lemma}}
        \renewcommand{\thetheorem}{\Alph{appendixc}.\arabic{theorem}}
        \renewcommand{\thedefinition}{\Alph{appendixc}.\arabic{definition}}
        \renewcommand{\thecorollary}{\Alph{appendixc}.\arabic{corollary}}
        \renewcommand{\theequation}{\Alph{appendixc}.\arabic{equation}}
        \noindent{\bf Appendix \theappendixc #1}\par\vspace{5pt}}
\newcommand{\subappendix}[1] {\vspace{14pt}
        \refstepcounter{subappendixc}
        \noindent{\bf Appendix \thesubappendixc. {\kern1pt \bfit #1}}
        \par\vspace{8pt}}
\newcommand{\subsubappendix}[1] {\vspace{14pt}
        \refstepcounter{subsubappendixc}
        \noindent{\rm Appendix \thesubsubappendixc. {\kern1pt \it #1}}
        \par\vspace{8pt}}

\newcommand{\textlineskip}{\baselineskip=13pt}
\newcommand{\smalllineskip}{\baselineskip=10pt}

\newcommand{\copyrightheading}[1]
        {\vspace*{-2.5cm}\smalllineskip{\flushleft
        {\footnotesize International Journal of Bifurcation and Chaos, #1}\\
        {\footnotesize \copyright\, World Scientific Publishing
         Company}\\
         }}

\newcommand{\publisher}[2]{{\begin{center}\tenrm\baselineskip=12pt
        Received #1\\
        Revised #2
        \end{center}
        }}

\def\abstracts#1#2{{
        \centering{\begin{minipage}{5.65in}\small\baselineskip=11pt
        \parindent=0pc #1\par 
        \parindent=2pc #2
        \end{minipage}}\par}}

\renewenvironment{thebibliography}[1]           %ALL CHANGES DD 13/3/92
        {\small\baselineskip=11pt
         \frenchspacing
         \begin{list}{\arabic{enumi}.}
        {\usecounter{enumi}\setlength{\parsep}{0pt}    
         \setlength{\leftmargin 12.7pt}{\rightmargin 0pt}%FOR 1--9 ITEMS
         \setlength{\itemsep}{0pt} \settowidth
        {\labelwidth}{#1.}\sloppy}}{\end{list}}

\newcounter{itemlistc}
\newcounter{romanlistc}
\newcounter{alphlistc}
\newcounter{arabiclistc}

\newcommand{\fcaption}[1]{
        \refstepcounter{figure}
        \setbox\@tempboxa = \hbox{\small Fig.~\thefigure. #1}
        \ifdim \wd\@tempboxa > 5in
           {\begin{center}
        \parbox{5in}{\small\baselineskip=11pt Fig.~\thefigure. #1}
            \end{center}}
        \else
             {\begin{center}
             {\small Fig.~\thefigure. #1}
              \end{center}}
        \fi}

\newcommand{\tcaption}[1]{
        \refstepcounter{table}
        \setbox\@tempboxa = \hbox{\small Table~\thetable. #1}
        \ifdim \wd\@tempboxa > 5in
           {\begin{center}
        \parbox{5in}{\small\baselineskip=11pt Table~\thetable. #1}
            \end{center}}
        \else
             {\begin{center}
             {\small Table~\thetable. #1}
              \end{center}}
        \fi}

\def\@citex[#1]#2{\if@filesw\immediate\write\@auxout
        {\string\citation{#2}}\fi
\def\@citea{}\@cite{\@for\@citeb:=#2\do
        {\@citea\def\@citea{,}\@ifundefined
        {b@\@citeb}{{\bf ?}\@warning
        {Citation `\@citeb' on page \thepage \space undefined}}
        {\csname b@\@citeb\endcsname}}}{#1}}

\newif\if@cghi
\def\cite{\@cghitrue\@ifnextchar [{\@tempswatrue
        \@citex}{\@tempswafalse\@citex[]}}
\def\citelow{\@cghifalse\@ifnextchar [{\@tempswatrue
        \@citex}{\@tempswafalse\@citex[]}}
\def\@cite#1#2{{$\null^{#1}$\if@tempswa\typeout
        {IJCGA warning: optional citation argument 
        ignored: `#2'} \fi}}

\def\pmb#1{\setbox0=\hbox{#1}
        \kern-.025em\copy0\kern-\wd0
        \kern.05em\copy0\kern-\wd0
        \kern-.025em\raise.0433em\box0}

\def\fnt#1#2{\footnotetext{\kern-.3em
        {$^{\mbox{\scriptsize #1}}$}{#2}}}

\font\tenrm=cmr10
\font\tenit=cmti10 

\font\bfit=cmbxti10 at 10pt

\font\eightit=cmti8

\def\itlatex{\tenit L\kern-.30em\raise.4ex\hbox{\eightit A}\kern-.14em 
T\kern-.1667em\lower.7ex\hbox{E}\kern-.125em X} 

\def\bsc{{\sc a\kern-7pt\sc a}}
\def\bflatex{\bf L\kern-.30em\raise.3ex\hbox{\bsc}\kern-.18em
T\kern-.1667em\lower.7ex\hbox{E}\kern-.125em X} 

\input amsfonts.sty
\input amssymb.sty

\begin{document}

\setlength{\textheight}{8.8truein}     %\VSIZE FOR 2ND PAGE  %8.8in

\thispagestyle{empty}

\markboth{V. Buchholtz \& T. P\"oschel}
{Adaptive Evolutionary Optimization of Team Work}

\textlineskip
\setcounter{page}{751}

\copyrightheading{Vol.~7, No.~3 (1997) 751-757} 

\vspace*{0.75truein}

\centerline{\large\bf ADAPTIVE EVOLUTIONARY OPTIMIZATION OF TEAM WORK}
\vspace*{0.28truein}

\centerline{VOLKHARD BUCHHOLTZ and THORSTEN P\"OSCHEL}
\vspace*{0.0215truein}
\centerline{\it Institut f\"ur Physik, Humboldt--Universit\"at zu Berlin,}
\centerline{Invalidenstr. 110, D--10115 Berlin, Germany}
\baselineskip=11pt
\vspace*{0.25truein}
\publisher{\quad  March 7, 1996}{\quad August 12, 1996}

\vspace*{0.25truein}
\abstracts{We discuss a new optimization strategy, which considerably improves
  the effectivity of evolutionary algorithms applied to a certain
  class of optimization problems. The basic principle is to solve
  first a simpler related problem, which is constructed by introducing
  additional degrees of freedom to the landscape. Starting from the
  solution in this simplified landscape we remove stepwise the added
  degrees of freedom. Our optimization strategy is demonstrated for a
  sample problem.}{}

\vspace*{0.35truein}\textlineskip

\begin{multicols}{2}
\section{Introduction}
\noindent
Evolutionary algorithms have been shown by many authors (e.g.~\cite{bsp})
to be suited for solving complex optimization problems, and there are a
highly developed theory of evolution processes~\cite{Ebeling} and lots
of recipes for successfully applying evolutionary
algorithms~\cite{Schwefel}. Particularly in combinatoric optimization
where many other standard techniques fail, considerable success has been achieved by using evolutionary algorithms. Obviously, each particular
optimization problem could be solved
more effectively (i.e. less computer time consuming) by a deterministic
algorithm, provided the algorithm is known. The main advantage of evolutionary algorithms is
their simplicity and universality. In most cases evolutionary algorithms can be parallelized in a trivial
way by distributing the individuals among a set of parallel
processors. This may be considered as another advantage.

One of the most well-known examples, which has drawn the attention of
scientists from various disciplines, is the Travelling Salesman Problem
(TSP). In its standard formulation, the TSP has been proven to be NP
complete~\cite{NP}, and many authors have tried to tackle this
problem using different types of evolutionary strategies as well as
other methods (e.g.~\cite{TSPEVO,TSPEVO1,TSPEVO1a,TSPEVO2,TSPEVO3,TSPEVO4}).

To apply an evolutionary algorithm one has to provide two essential ingredients:
First, one needs a fitness function which evaluates a solution. This
fitness function must be calculable very effectively (in terms of
computer time) since it has to be evaluated extremely often during the
calculation. Hence, the fitness function is required to be simple.
Second, one needs a mutation operator which takes into account the
topology of the fitness as a function of the parameters to be
optimized. If the mutation operator does not care about the topology
it happens that even a small mutation may lead to an extreme change in
the fitness, and hence, the evolution algorithm turns into stochastic
search.  Examples can be found in the literature where either the
first or the second precondition is violated and where the evolutionary
algorithm does not work effectively.

In the present paper we investigate an
evolutionary game where each individual is a set of points, i.e. a team,
which solves a well defined problem by cooperative behavior, i.e., by
``team work''. In the beginning neither the number of points per team
$N_{min}$ which is necessary to solve the problem nor the detailed
solution is known. Both have to be found during the evolutionary game.
Formally, one can consider the solution to be a subspace of the
configuration space of dimension $2~{N_{min}}$.

It will be shown that even if we knew the number of necessary points
it would be favourable to solve first a simpler problem, in which the teams
consist of more points than necessary, and then stepwise to increase
the complexity of the problem by reducing the number of points. Solving
such a hierarchy of problems and using the solution of a simple
problem as the initial condition for the next difficult problem can be
much more effective.

\section{Description of the Problem}
\label{sec:Problem}
\noindent
Assume we have a complicated shaped room with polygonal ground-plan.
The $M$ even walls of the room have to be illuminated completely using
a set of $N$ light bulbs. The questions which will be investigated
here are: {\em How many light bulbs are needed at minimum to illuminate
  the walls of the room, and where to place them?}

Although we cannot provide a proof for $NP$-completeness of the
problem, probably most people will agree that this problem is complex
in the general case. There is a short proof for the upper limit of bulbs
needed to illuminate a room bounded by $M$ even walls~\cite{stern}.
Provided that the room does not have columns one does not need more than
$N=\lfloor M/3\rfloor$ bulbs, where $\lfloor a\rfloor $ denotes the
integer of $a$. Figure~1 displays a room where one
needs indeed $N=\lfloor M/3 \rfloor$ bulbs. In many cases, however,
significantly less than $N=\lfloor M/3 \rfloor$ are necessary for
complete illumination. We want just to note that for the case that the
room has $S$ inner columns (of polygonal cross section) one claims (\cite{Rourke})
that $N=\lfloor (M+S)/3 \rfloor$ lamps are needed, where $M$ includes
the number of walls of the room and of the columns. So far, however, there is
no proof. \\[0.5cm]
\centerline{\psfig{file=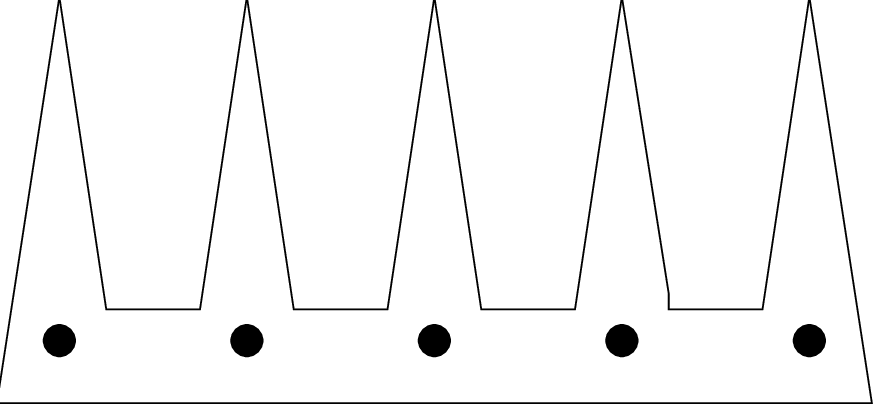,angle=270,width=5cm}}\\[0.4cm]
\begin{center}
\begin{minipage}{7.5cm}
Fig. 1. In the drawn case of a room with 15 walls we find the 
  worst case where $N=\lfloor 15/3\rfloor =5$ light bulbs will be
  needed to illuminate the walls completely. The dots show the
  positions of the bulbs.\\
\end{minipage}
\end{center}

A related problem is the so called {\em gallery watchman problem}\cite{Chvatal}. The
question here is how many (static) watchmen are needed to watch the
walls of a museum (e.g.~\cite{Stewart}), or in the dynamic formulation
what is the shortest path of a watchman to pass through all points of
interest. There are many formulations of the watchman problem and much
theoretic work has been done in this field
(e.g.~\cite{Rourke,GalleryWatchman,GalleryWatchman1,GalleryWatchman2,GalleryWatchman3}). A technical application of a one dimensional version of the dynamic watchman problem solved by an
evolutionary algorithm was recently investigated by
Heckman~\cite{Heckman}. A machine consisting of a number of pickers
assembled in a linear array across a conveyor belt was optimized to
pick up pieces which move on the conveyor. The algorithm had to decide
which of the pickers picks up the next coming piece.

In the case of our problem we assume that the room has no columns,
therefore the number $N_{min}$ of actually needed lamps is always less
than the third of the number of walls $N_{min} \le \lfloor M/3
\rfloor$.  The fitness $F$ of an individual, i.e., of a set of light
bulbs, is given by the illuminated area\footnote{Since the problem is
  two dimensional we may call the length of a certain part of the
  border of the room area.} of all walls
\begin{equation}
F_i = \frac{\sum\limits_{i=1}^M k_i}{\sum\limits_{i=1}^M l_i},
\label{fitness}
\end{equation}
where $k_i$ is the illuminated area of the $i$th wall and $l_i$ is its
total area. Hence we find for the fitness $F\in[0,1]$. During the
optimization we try to maximize $F$ by changing the positions of the
light bulbs. The solution is found when we have determined the
positions of the minimum of a number of bulbs to illuminate the room,
i.e. $F=1$. The function $F$ is embedded in the $2 N$ dimensional
space of the coordinates of the $N$ bulbs. It may have a complex
topology, in particular numerous discontinuities, local extrema and
flat plateaus.  Hence a simple gradient strategy would fail to find
the optimum and one has to chose a more sophisticated optimization
strategy. In light of
the generally used classification scheme introduced by
Schwefel~\cite{Schwefel1} our algorithm is of
$(\mu,\lambda)$--type.

Each individual $\alpha$, ($\alpha=1 \dots \mu$) in our evolutionary game
is a set of $N_{max} \ge N \ge N_{min}$ light bulbs, where $N_{max}
= \lfloor M/3 \rfloor$ and $N_{min}$ is the (unknown) minimal number
of bulbs which are needed to illuminate the room. The algorithm starts
with $\mu$ of such individuals, the positions $\vec{S}^\alpha_i$, $i=1
\dots N$ of the lamps are random. The optimization scheme reads as
follows:

\begin{enumerate}
\item The individuals are mutated by varying the position of each
  lamp
\begin{equation}
  {\vec{S}_i}^\alpha \rightarrow {\vec{S}_i}^\alpha + m_w \cdot
  \vec{A}_i^\alpha,
\end{equation}
with $i\in [1,N]$ and $\alpha \in [1,\mu]$. The components of the two dimensional vector $\vec{A}_i^\alpha$ are chosen equally distributed
from the interval $[0,1]$. The mutation step length $m_w$ is constant.
To avoid that the individuals persist in deep local maxima, with a
small frequency $P$ the new position of a lamp is completely random.
\item All individuals are rated by means of the fitness
  function~(\ref{fitness}).
  
\item If no set $\alpha\in [1\dots \mu]$ solves the problem, i.e.
  $F_\alpha \ne1$, the $\lambda$ sets of lamps (individuals) with the
  highest fitness values are copied and brought over to the next
  generation. The remaining $\mu - \lambda$ individuals die out. Steps
  1 to 3 are to be repeated until at least one of the individuals has
  the fitness $F_\alpha=1$.

\item If at least one individual has fitness $F=1$, the solution is
  found and we start to solve the next difficult problem, i.e., we try
  to solve the same optimization problem, now with each individual
  consisting of $N-1$ lamps only. The solution of the previous problem
  with $N$ lamps is used to initialize the new positions
  of $N-1$ lamps per individual: The winning individual with fitness
  $F=1$ is copied $\mu$ times and for each individual one randomly
  chosen lamp is removed. Then we start the new optimization run
  with $N-1$ bulbs beginning with item 1.
\end{enumerate}
This procedure is continued until no solution can be found
anymore. The last solution, found by the algorithm will be assumed to
be the solution of the optimization problem stated at the beginning of
this paragraph.\\[0.4cm]

\section{Results and discussion}
\noindent
The proposed algorithm was applied to illuminate a room ($M=82$)
with the ground-plan shown in Fig.~2. The solution
given by $N_{min}=10$ crosses was found during an evolutionary game of
$\mu=60$ individuals.  The optimization parameters were
$\lambda=5$, $P=10^{-3}$ and $m_w = \min(x_{max},y_{max}) \cdot
10^{-2}$ where $x_{max}$ and $y_{max}$ denote the maximum extent
of the room in the $x$- and $y$-directions respectively.\\[0.5cm]
  \centerline{\psfig{figure=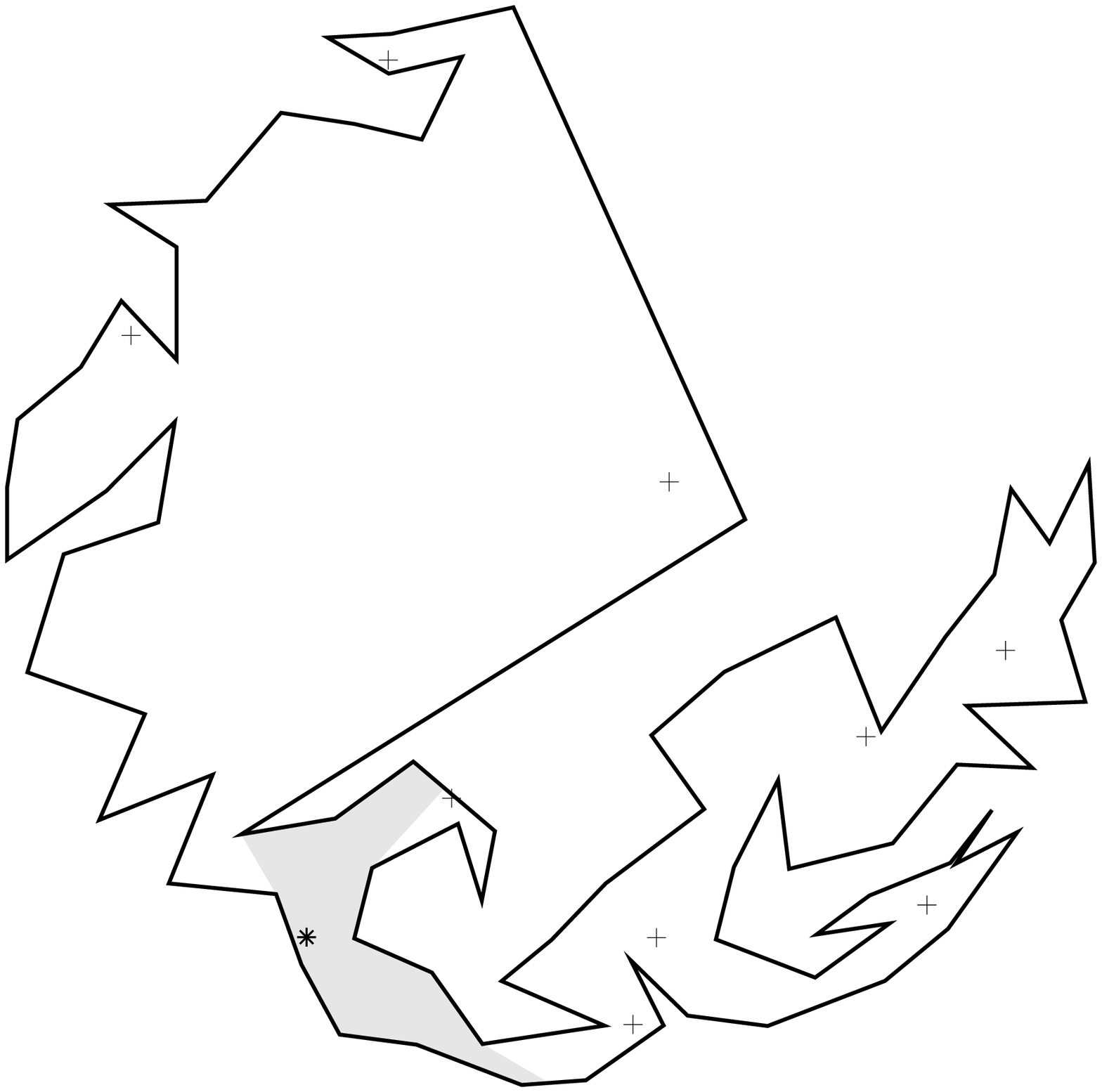,width=7.5cm,angle=0}}
%  \vspace{1cm}
  \begin{center}
    \begin{minipage}{7.5cm}
Fig. 2. Solution with $N_{min} = 10$ lamps for an
  arbitrary created room with $M=82$ walls. The position of the lamps
  are drawn with (+)- and (*)-symbols. The room is completely
  illuminated, one can easily check that each place of the wall can be
  connected by a straight line with at least one lamp without
  intersecting the wall. The shadowed area displays the section which
  is illuminated by the lamp which position is given by the
  (*)-symbol.
    \end{minipage}
  \end{center}
%\label{solution}
%\end{figure}

The parameters were chosen to give a satisfying efficiency of the optimization for a much simpler room. We did not further optimize these parameters, because the aim of the paper is to present a new optimization scheme but not to improve the efficiency of the well-known evolutionary algorithm. For the optimization of the parameters see e.g. ref.~\cite{Schwefel}.

In Fig.~3 the fitness $F_I$ of the fittest individual
$I$ is plotted versus the number of evolutionary steps. We start up with
random sets of $N=N_{min}=10$ lights. The fitness is not monotonously
increasing in time but there are long periods of stagnation
interrupted by rapid jumps in the fitness. (Note that the abscissa is
drawn in log scale.) This behavior, which seems to be typical for
evolutionary processes was observed by several authors for various
problems before, e.g.~\cite{Rechenberg}, and substantiated
theoretically (see~\cite{Ebeling}).
%\begin{figure}[ht]
\centerline{\psfig{figure=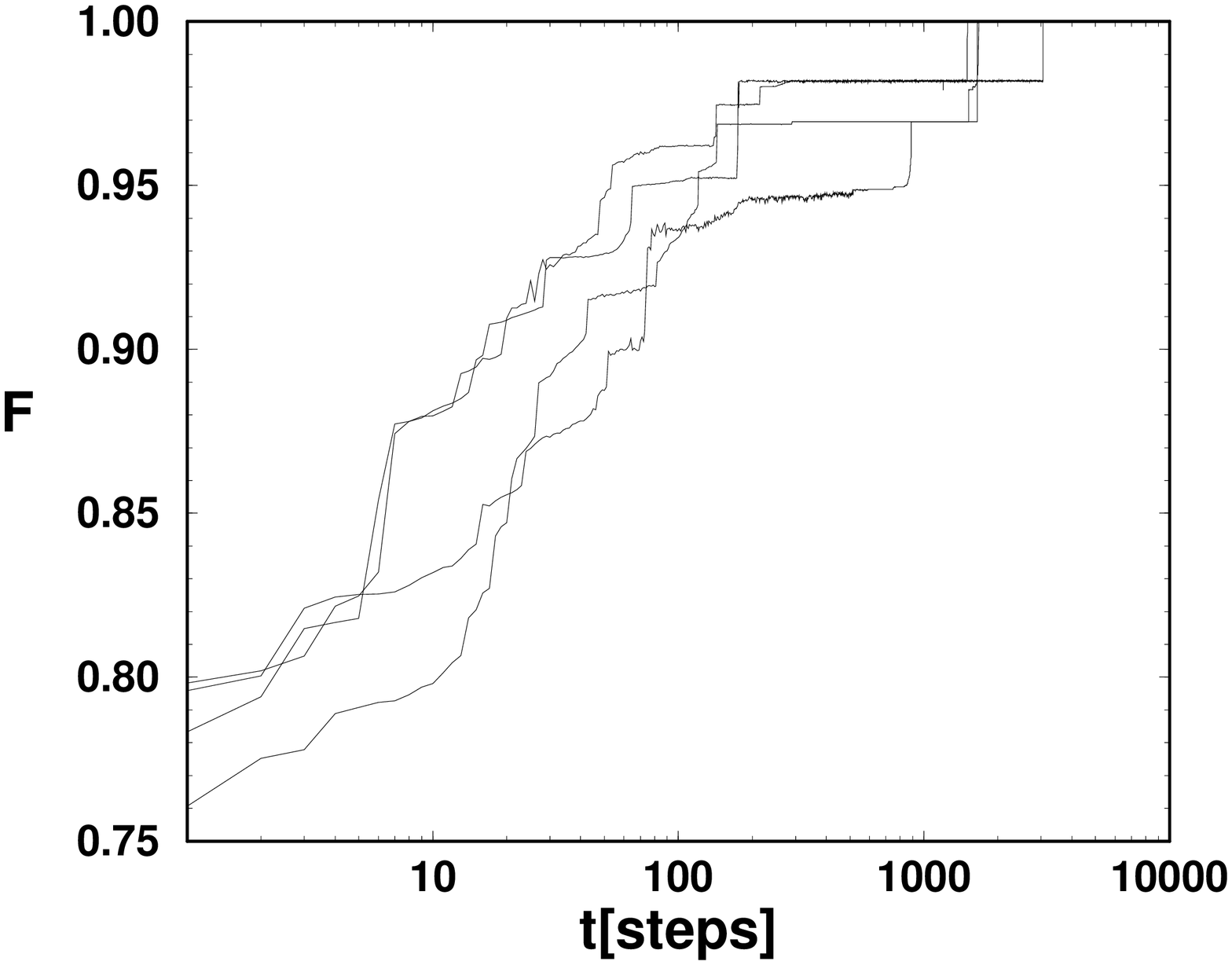,width=7.5cm,angle=270}}
\vspace{0.6cm}
\begin{center}
\begin{minipage}{7.5cm}
Fig. 3. Time evolution of the fitness $F$ of the best
  individual for four runs starting with different initial conditions.
  There are long periods of nearly constant fitness interrupted by
  short periods of rapid fitness increasing.\\
\end{minipage}
\end{center}
%\label{evolution}
%\end{figure}

Usually {\em a priori} we do not know the number $N_{min}$ of lamps
needed to illuminate the room and we start the optimization procedure
with $N_0>N_{min}$ lamps per individual. But even if we would know the
number it would be favourable to start with a larger amount of lamps
than needed. In the following we will show and explain, that the
algorithm for our optimization problem is up to about 10 times faster
for that case.

To ensure approximately the same amount of computer time for each
evolutionary step unaffected by the number of lamps each set consists of,
the number of individuals $\mu$ was chosen
\begin{equation}
\mu = \mu_{max} \cdot \frac{N_{min}}{N_0},
\end{equation}
where $\mu_{max}=60$ is the number of individuals for an
optimization starting with $N_0=N_{min}$ light bulbs
per individual. Hence, we can identify the number of evolutionary steps
with time when we assume that the computer time needed for each step
is mainly determined by the time required to calculate the
fitness, i.e., it is proportional to the total amount of light bulbs.

%\begin{figure}[ht]
\centerline{\psfig{figure=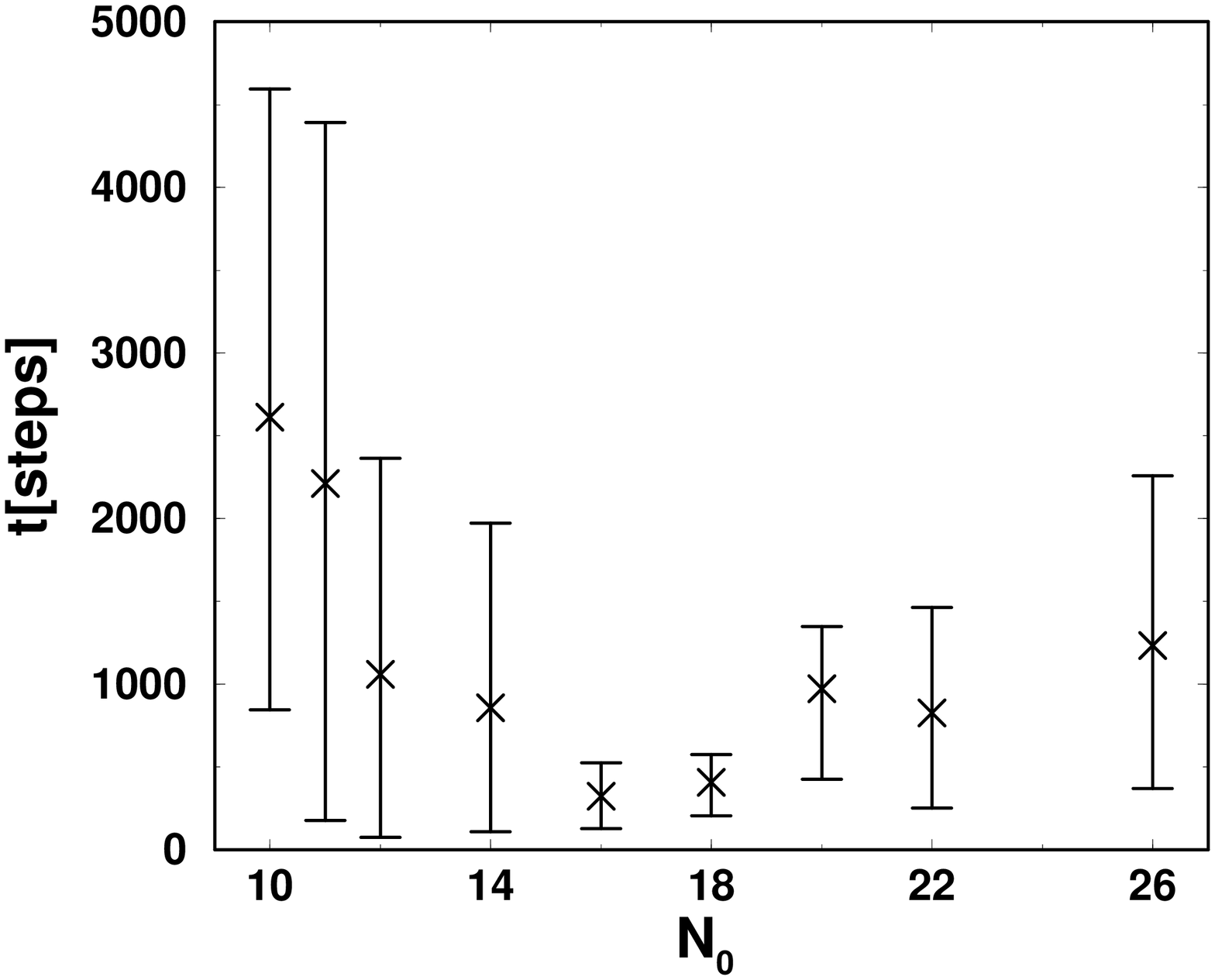,width=7.5cm,angle=270}}
\vspace{0.6cm}
\begin{center}
  \begin{minipage}{7.5cm}
Fig. 4. Number of optimization steps $t$ to find the final 
  solution versus the number of bulbs $N_0$ in the starting iteration.
  The computer time which is proportional to the number of iteration
  steps decreases with increasing $N_0$ and reaches a minimum at
  $N_0=16$.  For larger $N_0$ the value increases slowly.\\
  \end{minipage}
\end{center}
%\label{time_vs_N0}
%\end{figure}

Fig.~4 shows that the algorithm is in average about 10
times faster if one starts with sets consisting of $N_0=16$ lamps
instead of the required $N_{min}=10$. Each data point in this figure
is the mean value of the needed optimization steps of five runs with
different starting configurations. The data show that the time needed
to compute the solution decreases with increasing starting number of
lamps for $N_0<16$. When we start with larger sets $N_0>16$ the computer
time increases slowly. There are two mechanisms responsible for this
behavior:
\begin{enumerate}
\item Suppose we would do stochastic search, and suppose there would
  be a unique solution for the optimization problem. If we assume that
  $\tau$ is the time a single searcher needs to find one particular of
  the $N_{min}$ places then $N$ random independent searchers need the time
  \begin{equation}
    \tau^* = \frac{\tau}{N N_{min}}
  \end{equation}
  to find {\em any} of the places. The aspect of team work is included by assuming that a searcher
  who has found ``its place'' will not leave this place until the
  solution of the problem is found: It will survive the following
  evolutionary steps. Then we find for the time to find all places, i.e.
  to solve the optimization problem
\begin{equation}
  T(N) = \sum\limits_{i=0}^{N_{min}-1} \frac{\tau}{\left(N-i\right)\left(N_{min}-i\right)}~.
\label{startzeit}
\end{equation}
By replacing the sum by an integral we get the analytic solution
\begin{eqnarray}
  T(N)&\approx&   \tau \int\limits_{\frac{1}{2}}^{N_{min}-\frac{1}{2}} \frac{1}{(N-x)\left(N_{min}-x\right)}dx\nonumber \\
&& \!\!\!\!\!\!\!\!\!\!\!\!\!\!\!\!\!\!\!\!\! =\frac{\tau\ln\left(\frac{2N-1}{\left(2N_{min}-1\right)(2N-2N_{min}+1)}\right)}{N_{min}-N}~.
\end{eqnarray}
Obviously $T(N)-T(N+1)$ is a positive number, i.e. the solution will
be found quicker when starting with $N+1$ lamps instead of $N$.

%\begin{figure}[ht]
\centerline{\psfig{figure=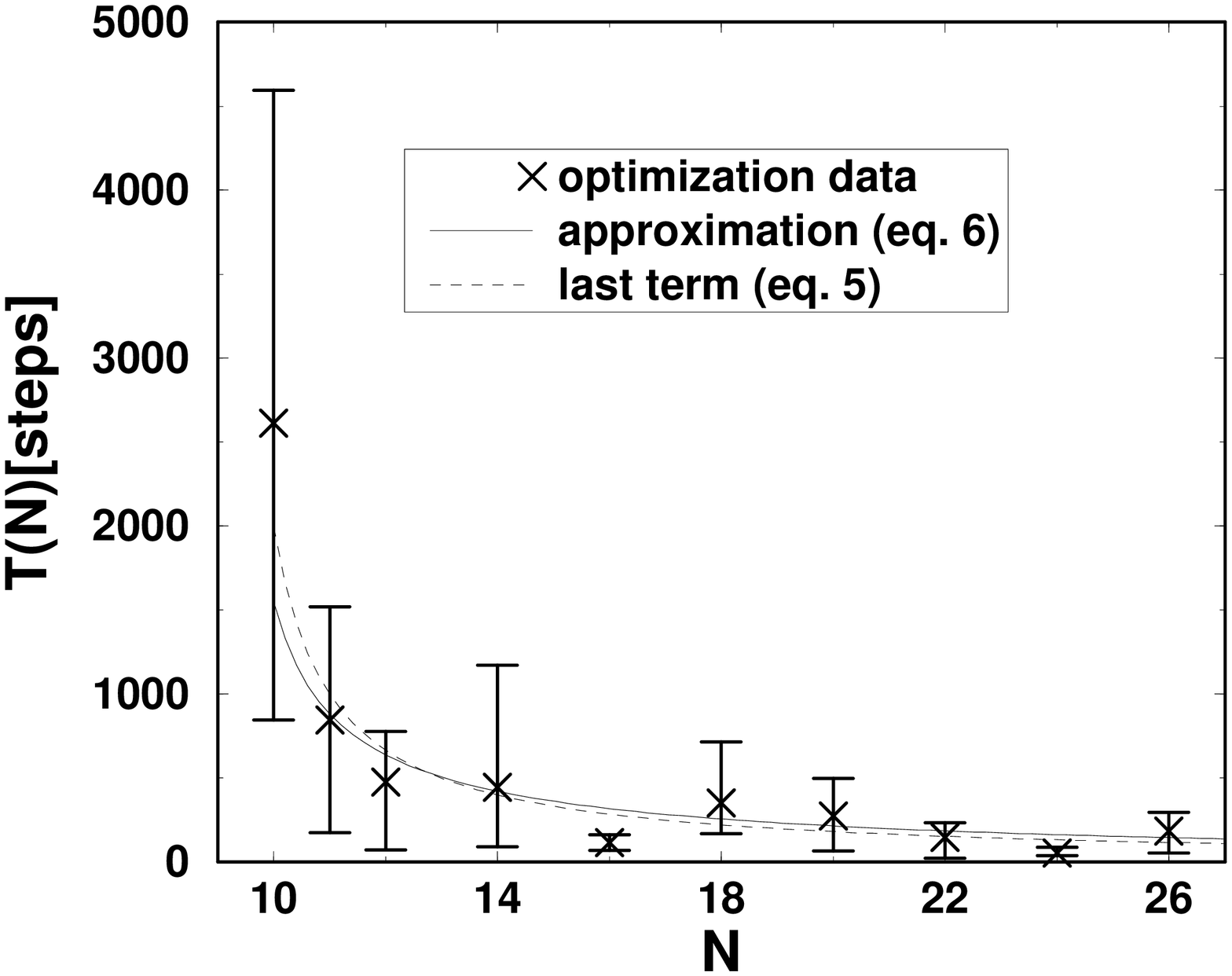,width=7.5cm,angle=270}}
%\vspace{1cm}
\begin{center}
  \begin{minipage}{7.5cm}
Fig. 5. Number of optimization steps $T_N$ versus the number of lamps
  $N$. The averaged data over five runs of the evolution algorithm are
  drawn by crosses and error bars. The full line and the dashed line
  show the estimates due to eq.~(\ref{startzeit}) and the last term of
  the sum in eq.~(\ref{startzeit}) respectively. The time $\tau$ has
  been assumed to be a fit parameter.
  \end{minipage}
\end{center}
%\label{grund1}
%\end{figure}

Apart from this simple estimate we realize that typically the
``attractor regions'' for the positions of the bulbs do not have the
same size but their sizes differ significantly. Usually the positions
with the smallest ``attractors'' are found at the end of the
optimization procedure, and the time is mainly determined by the last
term of the sum in eq.~(\ref{startzeit}). To provide better estimate
one needs knowledge about the fitness landscape of the problem. For
very simple problems it has been shown that one can conclude the
properties of convergence based on knowledge od statistical properties
of the fitness landscape~\cite{Asselmeyer}.

Fig.~5 shows the results from the optimization as well as
the discussed estimates.\\[0.3cm]

%\begin{figure}[ht]
\centerline{~~~~~~~~~~\psfig{figure=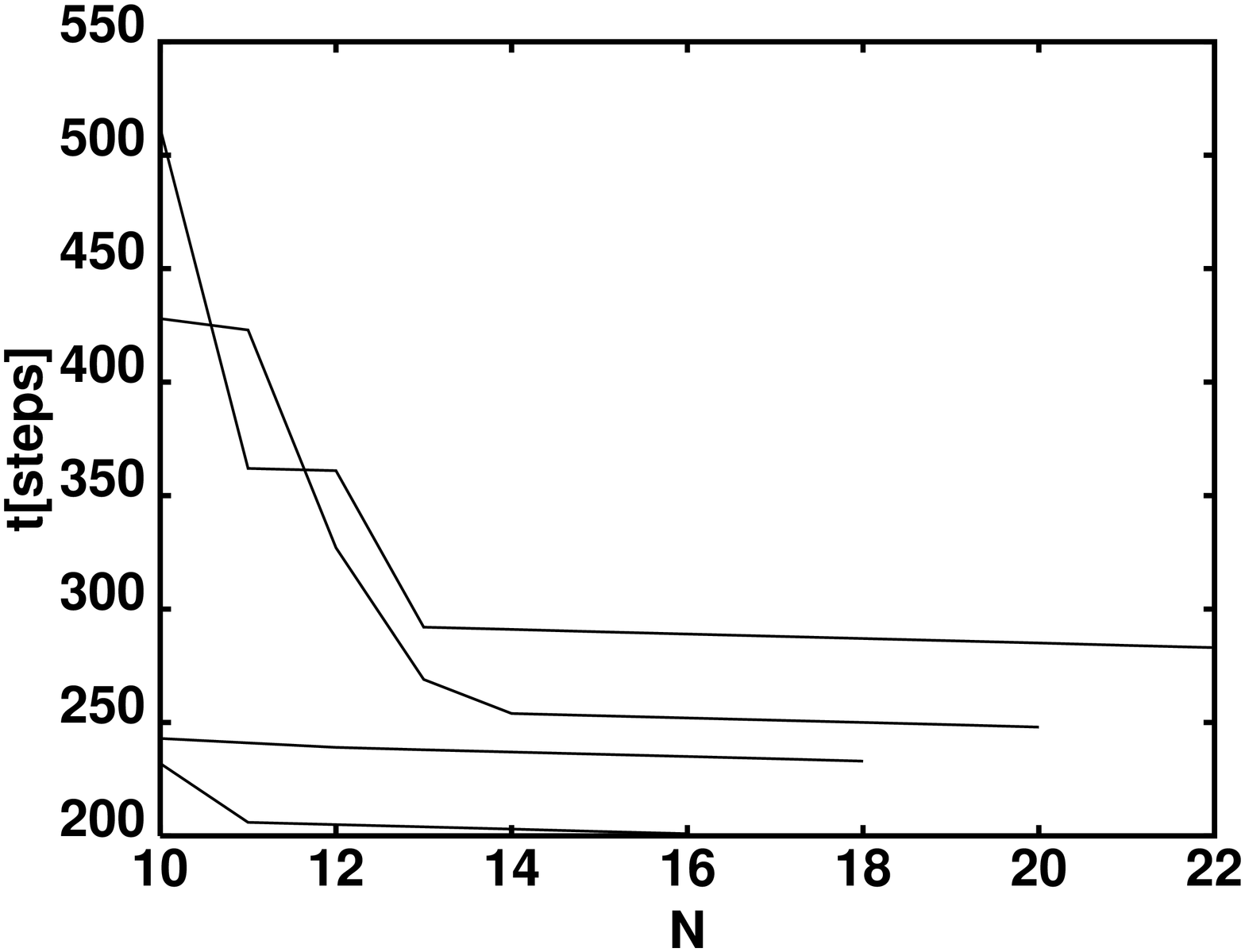,width=7cm,angle=270}}
%\vspace{0.6cm}
\begin{center}
  \begin{minipage}{7.5cm}
Fig. 6. Sample optimization runs. After finding a solution for $N_0>N_{min}$ 
  lamps the algorithm finds quickly the solution for smaller $N$. In accordance with fig.~4 the algorithm becomes slightly less effective when starting with larger $N_0$ (Explanation see text).
  \end{minipage}
\end{center}
%\label{grund2}
%\end{figure}

\item Fig.~6 shows the progress of the optimization procedure
  for $N_0=\{16,18,20,22\}$ lights in the starting configuration. One
  should read the figure from right to left: the ordinate shows the
  number of evolution cycles which have been necessary to solve the
  problem for $N$ lamps. In the case of the run started with
  $N_0=18$ light bulbs the solution for $N=N_{min}$ was found
  almost exclusively by removing lights.
  
  Fig.~4 shows that the optimization time rises
  slightly for large $N_0$. Comparing the solutions of one
  optimization run for different $N$ we note that for $N_0
  \stackrel{<}{\sim} 18$ the initial solution (for $N=N_0$) is
  frequently very close to the final solution (for $N=N_{min}$).
  Hence, the solution for $N-1$ lamps can be found from the solution
  for $N$ lamps just by deleting one of the lamps and performing few
  evolution steps (see Fig.~6).
  
  For $N_0 \stackrel{>}{\sim} 18$ the initial solution could differ
  very much from the final solution, and also the solutions of the
  successive problems for $N_0$, $N_0-1 \dots ,N_{min}$ differ from
  each other. The solutions are not really adapted to the geometry of
  the room, but they are more the results of a random search.
  Therefore the system does not take too much advantage from knowing
  the solution for $N$ lamps solving the problem for $N-1$ lamps.
  Hence the calculation time in Fig.~4 rises slightly
  for larger values of $N$. Nevertheless the algorithm is still much
  faster compared with the time needed for the solution starting with
  $N=N_{min}$ bulbs.

  The situation is quite similar to a neural network applied to a
  pattern recognition problem: if it has too many neurons the network
  just stores the patterns instead of finding characteristic features~(e.g.~\cite{NN}).
  If this network is applied to an unknown pattern it fails since none of
  the stored patterns has enough overlap with the unknown pattern. A
  network with less neurons might be able to solve the problem since
  it checks whether the unknown pattern reveals characteristic
  features.
\end{enumerate}

\section{Conclusion}
We have shown that for the investigated problem it is favourable to
start the optimization with more points than necessary $N_0 >
N_{min}$. First we find a solution, i.e. the positions of the points
for a higher number of points, and then we stepwise decrease $N$ while
using the solution for $N$ as initial condition for finding the
solution for $N-1$ points. It turns out that the procedure to solve
the chain of problems for $N_0, N_0-1, N_0-2, \dots , N_{min}$ is up
to about ten times faster then to solve the problem for $N_{min}$
directly. Formally we first solve the optimization problem in a high
dimensional space of dimension $2 N_0$. Once the solution is found we
stepwise reduce the dimensionality and end up with dimension
$2 N_{min}$. 

The theoretical basis of the optimization speedup in higher dimension
is provided by Morse theory~\cite{Morse}. Qualitatively Morse Lemma
says that under rather mild assumptions the fraction of saddle points
of a function of its critical points rises with dimension. Hence the
relation of extrema and saddle points decreases when increasing the
dimension. The idea of the lemma becomes clear in one and two
dimensions: to reach the optimum of a one dimensional function one has
to ``walk through'' all local extrema in between the starting point and
the global extremum. In two dimensions in many cases one can ``walk
around'' local extrema and hence one avoids time consuming escape
procedures. Similar in higher dimension: the higher the dimension the
more bypasses do exist to reach a certain point without getting stuck
in local extrema.

We assume that this behavior is typical for a certain class of
optimization problems. The main characteristic of this class is that
the problem can be transformed into another problem containing a
parameter $N$ where the solution of the original problem with the
parameter being $N^*$ is contained in the set of solutions of the
 easier to solve problems with $N\ne N^*$. Obviously the entire class of coverage problems belongs to this class. Further investigations will be
necessary to substantiate this hypothesis. For the problem of the
travelling salesman, for instance, the easier task could be to find a
minimal length where some cities ($N$) are allowed to be visited twice,
followed by a successively reduction of the number of those cities.
\\[0.3cm]

The authors thank T.~Asselmeyer, W.~Ebeling, H.~Herzel, G.~Rudolph and
L.~Schimansky-Geier for stimulating discussions. V.B. thanks project
EVOALG (BMBF) for financial support.

\end{multicols}

\end{document}